\documentclass[aps,prl,twocolumn,superscriptaddress,showpacs]{revtex4} 
\usepackage{graphicx}
\usepackage{dcolumn}
\usepackage{bm}
\usepackage[abs]{overpic}
\usepackage{multirow}
\usepackage{booktabs}

\begin{document}
\newlength{\tmpLength}

\def\vmet{\mbox{$\protect \raisebox{0.3ex}{$\not$}\vec{E}_\mathrm{T}$}}
\def\met{\mbox{$\protect \raisebox{0.3ex}{$\not$}E_\mathrm{T}$}}
\def\mht{\mbox{$\protect \raisebox{0.3ex}{$\not$}H_\mathrm{T}$}}
\def\vmpt{\mbox{$\protect \raisebox{0.3ex}{$\not$}\vec{p}_\mathrm{T}^{\text{tr}}$}}
\def\mpt{\mbox{$\protect \raisebox{0.3ex}{$\not$}p_\mathrm{T}^{\text{tr}}$}}
\def\Pt{{p_\mathrm{T}^{\text{tr}}}}
\def\PtRelDef{\mbox{\ensuremath{Z[j] \equiv \sum p_\mathrm{T}^{\text{tr}, j} / p_\mathrm{T}^{j}}}}
\def\JetPtRelOne{\ensuremath{Z[j_1]}}
\def\JetPtRelTwo{\ensuremath{Z[j_2]}}

\def\met{\mbox{${\not\!\!E_T}$}}
\def\vmet{\mbox{${\not\!\!\vec{E}_T}$}}
\def\mpt{\mbox{${\not\!\!p^{\text{tr}}_T}$}}
\def\vmpt{\mbox{${\not\!\!\vec{p}^{\text{tr}}_T}$}}
\def\mht{\mbox{${\not\!\!H_T}$}}

\def\invfb{\,fb$^{-1}$}
\def\gevcc{\,\,\mathrm{GeV}/c^2} 
\def\gevc{\,\,\mathrm{GeV}/c} 
\def\gev{\,\,\mathrm{GeV}} 


\title{Search for the standard model Higgs boson decaying to a $b\bar{b}$ pair in events with no charged leptons and large missing transverse energy 
using the full CDF data set}

\affiliation{Institute of Physics, Academia Sinica, Taipei, Taiwan 11529, Republic of China}
\affiliation{Argonne National Laboratory, Argonne, Illinois 60439, USA}
\affiliation{University of Athens, 157 71 Athens, Greece}
\affiliation{Institut de Fisica d'Altes Energies, ICREA, Universitat Autonoma de Barcelona, E-08193, Bellaterra (Barcelona), Spain}
\affiliation{Baylor University, Waco, Texas 76798, USA}
\affiliation{Istituto Nazionale di Fisica Nucleare Bologna, $^{ee}$University of Bologna, I-40127 Bologna, Italy}
\affiliation{University of California, Davis, Davis, California 95616, USA}
\affiliation{University of California, Los Angeles, Los Angeles, California 90024, USA}
\affiliation{Instituto de Fisica de Cantabria, CSIC-University of Cantabria, 39005 Santander, Spain}
\affiliation{Carnegie Mellon University, Pittsburgh, Pennsylvania 15213, USA}
\affiliation{Enrico Fermi Institute, University of Chicago, Chicago, Illinois 60637, USA}
\affiliation{Comenius University, 842 48 Bratislava, Slovakia; Institute of Experimental Physics, 040 01 Kosice, Slovakia}
\affiliation{Joint Institute for Nuclear Research, RU-141980 Dubna, Russia}
\affiliation{Duke University, Durham, North Carolina 27708, USA}
\affiliation{Fermi National Accelerator Laboratory, Batavia, Illinois 60510, USA}
\affiliation{University of Florida, Gainesville, Florida 32611, USA}
\affiliation{Laboratori Nazionali di Frascati, Istituto Nazionale di Fisica Nucleare, I-00044 Frascati, Italy}
\affiliation{University of Geneva, CH-1211 Geneva 4, Switzerland}
\affiliation{Glasgow University, Glasgow G12 8QQ, United Kingdom}
\affiliation{Harvard University, Cambridge, Massachusetts 02138, USA}
\affiliation{Division of High Energy Physics, Department of Physics, University of Helsinki and Helsinki Institute of Physics, FIN-00014, Helsinki, Finland}
\affiliation{University of Illinois, Urbana, Illinois 61801, USA}
\affiliation{The Johns Hopkins University, Baltimore, Maryland 21218, USA}
\affiliation{Institut f\"{u}r Experimentelle Kernphysik, Karlsruhe Institute of Technology, D-76131 Karlsruhe, Germany}
\affiliation{Center for High Energy Physics: Kyungpook National University, Daegu 702-701, Korea; Seoul National University, Seoul 151-742, Korea; Sungkyunkwan University, Suwon 440-746, Korea; Korea Institute of Science and Technology Information, Daejeon 305-806, Korea; Chonnam National University, Gwangju 500-757, Korea; Chonbuk National University, Jeonju 561-756, Korea}
\affiliation{Ernest Orlando Lawrence Berkeley National Laboratory, Berkeley, California 94720, USA}
\affiliation{University of Liverpool, Liverpool L69 7ZE, United Kingdom}
\affiliation{University College London, London WC1E 6BT, United Kingdom}
\affiliation{Centro de Investigaciones Energeticas Medioambientales y Tecnologicas, E-28040 Madrid, Spain}
\affiliation{Massachusetts Institute of Technology, Cambridge, Massachusetts 02139, USA}
\affiliation{Institute of Particle Physics: McGill University, Montr\'{e}al, Qu\'{e}bec, Canada H3A~2T8; Simon Fraser University, Burnaby, British Columbia, Canada V5A~1S6; University of Toronto, Toronto, Ontario, Canada M5S~1A7; and TRIUMF, Vancouver, British Columbia, Canada V6T~2A3}
\affiliation{University of Michigan, Ann Arbor, Michigan 48109, USA}
\affiliation{Michigan State University, East Lansing, Michigan 48824, USA}
\affiliation{Institution for Theoretical and Experimental Physics, ITEP, Moscow 117259, Russia}
\affiliation{University of New Mexico, Albuquerque, New Mexico 87131, USA}
\affiliation{The Ohio State University, Columbus, Ohio 43210, USA}
\affiliation{Okayama University, Okayama 700-8530, Japan}
\affiliation{Osaka City University, Osaka 588, Japan}
\affiliation{University of Oxford, Oxford OX1 3RH, United Kingdom}
\affiliation{Istituto Nazionale di Fisica Nucleare, Sezione di Padova-Trento, $^{ff}$University of Padova, I-35131 Padova, Italy}
\affiliation{University of Pennsylvania, Philadelphia, Pennsylvania 19104, USA}
\affiliation{Istituto Nazionale di Fisica Nucleare Pisa, $^{gg}$University of Pisa, $^{hh}$University of Siena and $^{ii}$Scuola Normale Superiore, I-56127 Pisa, Italy}
\affiliation{University of Pittsburgh, Pittsburgh, Pennsylvania 15260, USA}
\affiliation{Purdue University, West Lafayette, Indiana 47907, USA}
\affiliation{University of Rochester, Rochester, New York 14627, USA}
\affiliation{The Rockefeller University, New York, New York 10065, USA}
\affiliation{Istituto Nazionale di Fisica Nucleare, Sezione di Roma 1, $^{jj}$Sapienza Universit\`{a} di Roma, I-00185 Roma, Italy}
\affiliation{Rutgers University, Piscataway, New Jersey 08855, USA}
\affiliation{Texas A\&M University, College Station, Texas 77843, USA}
\affiliation{Istituto Nazionale di Fisica Nucleare Trieste/Udine, I-34100 Trieste, $^{kk}$University of Udine, I-33100 Udine, Italy}
\affiliation{University of Tsukuba, Tsukuba, Ibaraki 305, Japan}
\affiliation{Tufts University, Medford, Massachusetts 02155, USA}
\affiliation{University of Virginia, Charlottesville, Virginia 22906, USA}
\affiliation{Waseda University, Tokyo 169, Japan}
\affiliation{Wayne State University, Detroit, Michigan 48201, USA}
\affiliation{University of Wisconsin, Madison, Wisconsin 53706, USA}
\affiliation{Yale University, New Haven, Connecticut 06520, USA}

\author{T.~Aaltonen}
\affiliation{Division of High Energy Physics, Department of Physics, University of Helsinki and Helsinki Institute of Physics, FIN-00014, Helsinki, Finland}
\author{B.~\'{A}lvarez~Gonz\'{a}lez$^z$}
\affiliation{Instituto de Fisica de Cantabria, CSIC-University of Cantabria, 39005 Santander, Spain}
\author{S.~Amerio}
\affiliation{Istituto Nazionale di Fisica Nucleare, Sezione di Padova-Trento, $^{ff}$University of Padova, I-35131 Padova, Italy}
\author{D.~Amidei}
\affiliation{University of Michigan, Ann Arbor, Michigan 48109, USA}
\author{A.~Anastassov$^x$}
\affiliation{Fermi National Accelerator Laboratory, Batavia, Illinois 60510, USA}
\author{A.~Annovi}
\affiliation{Laboratori Nazionali di Frascati, Istituto Nazionale di Fisica Nucleare, I-00044 Frascati, Italy}
\author{J.~Antos}
\affiliation{Comenius University, 842 48 Bratislava, Slovakia; Institute of Experimental Physics, 040 01 Kosice, Slovakia}
\author{G.~Apollinari}
\affiliation{Fermi National Accelerator Laboratory, Batavia, Illinois 60510, USA}
\author{J.A.~Appel}
\affiliation{Fermi National Accelerator Laboratory, Batavia, Illinois 60510, USA}
\author{T.~Arisawa}
\affiliation{Waseda University, Tokyo 169, Japan}
% ...

\collaboration{CDF Collaboration\footnote[2]{With visitors from
$^a$Istituto Nazionale di Fisica Nucleare, Sezione di Cagliari, 09042 Monserrato (Cagliari), Italy,
$^b$University of California Irvine, Irvine, CA 92697, USA,
$^c$University of California Santa Barbara, Santa Barbara, CA 93106, USA,
$^d$University of California Santa Cruz, Santa Cruz, CA 95064, USA,
$^e$Institute of Physics, Academy of Sciences of the Czech Republic, Czech Republic,
$^f$CERN, CH-1211 Geneva, Switzerland,
$^g$Cornell University, Ithaca, NY 14853, USA,
$^h$University of Cyprus, Nicosia CY-1678, Cyprus,
$^i$Office of Science, U.S. Department of Energy, Washington, DC 20585, USA,
$^j$University College Dublin, Dublin 4, Ireland,
$^k$ETH, 8092 Zurich, Switzerland,
$^l$University of Fukui, Fukui City, Fukui Prefecture, Japan 910-0017,
$^m$Universidad Iberoamericana, Mexico D.F., Mexico,
$^n$University of Iowa, Iowa City, IA 52242, USA,
$^o$Kinki University, Higashi-Osaka City, Japan 577-8502,
$^p$Kansas State University, Manhattan, KS 66506, USA,
$^q$Ewha Womans University, Seoul, 120-750, Korea,
$^r$University of Manchester, Manchester M13 9PL, United Kingdom,
$^s$Queen Mary, University of London, London, E1 4NS, United Kingdom,
$^t$University of Melbourne, Victoria 3010, Australia,
$^u$Muons, Inc., Batavia, IL 60510, USA,
$^v$Nagasaki Institute of Applied Science, Nagasaki, Japan,
$^w$National Research Nuclear University, Moscow, Russia,
$^x$Northwestern University, Evanston, IL 60208, USA,
$^y$University of Notre Dame, Notre Dame, IN 46556, USA,
$^z$Universidad de Oviedo, E-33007 Oviedo, Spain,
$^{aa}$CNRS-IN2P3, Paris, F-75205 France,
$^{bb}$Texas Tech University, Lubbock, TX 79609, USA,
$^{cc}$Universidad Tecnica Federico Santa Maria, 110v Valparaiso, Chile,
$^{dd}$Yarmouk University, Irbid 211-63, Jordan.
}}
\noaffiliation

\date{\today}

\begin{abstract}

We report on a search for the standard model Higgs boson produced in association with a vector boson in the full data set of proton-antiproton collisions at $\sqrt{s} = 1.96$ TeV recorded by the CDF~II detector at the Tevatron, corresponding to an integrated luminosity of 9.45\invfb.
We consider events having no identified charged lepton, a transverse energy imbalance, and two or three jets, of which at least one is consistent with originating from the decay of a $b$ quark.
We place 95\% credibility level upper limits on the production cross section times standard model branching fraction for several mass hypotheses between $90$ and $150\gevcc$.
For a Higgs boson mass of $125\gevcc$, the observed (expected) limit is $6.7$ ($3.6$) times the standard model prediction.
\end{abstract}


\maketitle

In the standard model (SM)~\cite{sm}, the mechanism responsible for spontaneous electroweak symmetry breaking gives mass to the $W$ and $Z$ bosons~\cite{higgs}. The Higgs boson ($H$) represents the remaining degree of freedom after the symmetry is broken and also allows fermions to acquire mass through Yukawa couplings. 
The SM does not predict the mass of the Higgs boson, $m_H$, but the combination of precision electroweak measurements~\cite{elweak}, including recent top quark and $W$ boson mass measurements from the Tevatron~\cite{topmass,wmass}, constrains $m_H<152\gevcc$ at the 95\% confidence level.    Direct searches at LEP2~\cite{sm-lep}, the Tevatron~\cite{tevcomb2012}, and the LHC~\cite{sm-lhc} exclude all possible masses of the SM Higgs boson at the 95\% confidence level or the 95\% credibility level (C.L.), except within the ranges 116.6 -- 119.4$\gevcc$ and 122.1 -- 127$\gevcc$.  A SM Higgs boson in this mass range would be produced in the $\sqrt{s}=1.96$ TeV $p{\bar{p}}$ collisions of the Tevatron, and would have a branching fraction to $b \bar{b}$ greater than 50\%~\cite{vhtheory, stange, lhcdifferential}.
In these currently allowed regions, $H\to b\bar{b}$ is the dominant decay mode, but large QCD multi-jet backgrounds overwhelm searches in the exclusive $b\bar{b}$ final state. Searches for $H$ produced in association with a vector boson, {\it VH} ($V=W$ or $Z$), where the vector boson decays leptonically, access final states with 
significantly higher signal to background ratios than those resulting from $gg\to H\to b\bar{b}$.

This Letter presents a search for {\it VH} production in events with missing transverse energy ($\met$)~\cite{coord} -- a signature of neutrinos escaping detection -- and $b$ jets in a data set corresponding to an integrated luminosity of 9.45\invfb\ collected using the CDF~II detector at the Fermilab Tevatron. 
This analysis considers $Z(\to\nu\bar{\nu})H$ production, where the neutrinos ($\nu$) escape detection, or $Z(\to\ell^+\ell^-)H$ when neither charged lepton is identified or they are reconstructed as jets. We are also sensitive to {\it WH} events where $W\to e\nu$ or $W\to\tau\nu$ and the charged lepton is reconstructed as a jet, or where it is not identified.
By building upon techniques used for the observation of single-top-quark production~\cite{cdfmetbb_singletop}, we significantly increase the signal acceptance with respect to previous Tevatron searches in this final state~\cite{Aaltonen:2009jg,dZeroMETbb}.

CDF~II is a multi-purpose collider detector described in Ref.~\cite{CDFII}. A three-level online selection system (trigger) is used to select events for analysis. Events are selected via boolean OR of two trigger paths~\cite{triggerPath} requiring either the presence of large $\met$, or large $\met$ and two jets~\cite{cdfjets}.  The efficiency associated to this selection is obtained from data and is applied to the Monte-Carlo (MC) simulated samples to reproduce the inefficiencies present in the data.  
The parametrization of the trigger efficiency~\cite{kaThesis} significantly improves the modeling of the trigger turn-on outside the fully efficient region, as verified using data control samples. This allows significantly relaxed pre-selection requirements compared to that of  Ref.~\cite{Aaltonen:2009jg}. 
The parametrization is done using a neural network (NN)~\cite{neuralnets} trained from the following inputs: the $\met$ in the event, its azimuth ($\varphi(\vmet)$), three variables characterizing the $i$th jet ($j_i$) in the event -- $E_T(j_i)$, $\eta(j_i)$ and $\varphi(j_i)$ -- and the $\eta$-$\phi$ separation of the jets $\Delta R = \sqrt{\Delta\varphi^2 + \Delta\eta^2}$~\cite{coord}.  
We thus have 9 (14) input variables for events with two (three) jets. 
We use a muon-triggered sample to define the nominal parametrization and derive the trigger systematic uncertainty from a parametrization of an inclusive jet sample with at least one jet with $E_T>50\gev$. The efficiency ranges from 0.40 for events having $\met = 35~\gev$ to 1.0 for events with $\met>80~\gev$. 

We reconstruct jets from energy depositions in the calorimeter towers using a jet clustering cone algorithm~\cite{cdfjets} with a cone of radius $\Delta R = 0.4$. In addition to the standard jet-energy corrections used by CDF~\cite{cdfjets}, we adjust the energy of the jets according to the measured momentum of the charged particle tracks within the jet cone~\cite{H1}. We further improve the energy determination using a NN approach to estimate the energies of the initiating quarks.   
The direction and magnitude of the $\vmet$ are then recomputed.  These jet reconstruction methods improve both the signal acceptance and the relative resolution of the reconstructed invariant mass of the Higgs boson candidate by $\sim15\%$.  We reject events with an identified $e$ or $\mu$  
to maintain statistical independence from other CDF analyses searching for the SM Higgs boson~\cite{llbb,lvbb}. 

After the events are reconstructed, the following pre-selection requirements are made: we select events with \mbox{$\met>35~\gev$} and two or three jets satisfying \mbox{$E_T>15\gev$} and $|\eta|<2.4$, thus accepting events where partons provide an additional jet candidate, or a lepton ($e$ or $\tau$) is reconstructed as a jet.  The two most energetic jets, $j_1$ and $j_2$, are required to have reconstructed transverse energies of at least $25$ and $20~\gev$ respectively, satisfy $\left|\eta(j_i)\right|<2$, be separated 
by $\Delta R(j_1,j_2)>0.8$, and at least one of these two jets must satisfy $|\eta(j_i)|<0.9$.
This selection is relaxed with respect to Ref.~\cite{Aaltonen:2009jg}, and increases the signal acceptance by a factor of $1.4$.
The cost of this increased signal acceptance is a tenfold increase of the background acceptance.  One of the leading sources of significant \met\ in  QCD (quantum chromodynamics) production of multi-jet events (QCD MJ) arises from the mismeasurement of jet energies.  Neutrinos from semileptonic $b$ decays can also produce significant \met{} in QCD MJ events.   In both of those cases, the \vmet\ is often aligned with $\vec{E}_\mathrm{T}^{j_2}$, and such events are rejected by requiring \mbox{$\Delta\varphi(\vmet,\vec{E}_\mathrm{T}^{j_1}) \geq 1.5$} and \mbox{$\Delta\varphi(\vmet,\vec{E}_\mathrm{T}^{j_{2,3}}) \geq 0.4$}. %
This reduces the backgrounds by a factor of three, while retaining 90\% of the signal.  
The large backgrounds from light-flavor jet production originating from $u$, $d$, or $s$ quarks or gluons are reduced by identifying (tagging) jets consistent with the decay of $b$ quarks; $c$ quarks are not explicitly identified.

We use two algorithms to tag $b$-quark jets, SECVTX~\cite{secvtx}, which attempts to reconstruct the secondary vertex from the $b$ decay (displaced from the interaction point because hadrons containing $b$ or $c$ quarks can travel a few millimeters in the detector before decaying), and JETPROB~\cite{jetprob}, which determines for each jet the probability that the tracks within the jet are consistent with originating from the primary vertex. 
We operate SECVTX (JETPROB) at about 40\% (50\%) efficiency, yielding a rate of light-flavor jets mistakenly identified as $b$ jets (mistags) of about 1\% (5\%). 
We exploit the different purities of the selected multi-tagged events by considering independent tagging categories separately and later combining results. 
We require that one of the leading jets be tagged by SECVTX and the other be tagged either by SECVTX (SS) or JETPROB (SJ), or be untagged (1S). The tagging process reduces the backgrounds by two orders of magnitude while retaining about $50\%$ of the signal. Events satisfying the aforementioned criteria comprise the {\it preselection} sample.
 The signal-to-background ratio ($S/B$) in this sample is estimated to be \mbox{$S/B\sim 1/400$} in the SS tagging category  for $m_H=125~\gevcc$,  compared to less than $10^{-5}$ for the full sample of triggered events. The relative fraction of events with $Z\to\nu\bar{\nu}$, $Z\to\ell^+\ell^-$, and $W\to\ell\nu$ is respectively 47, 3 and 50 percent; of the latter, the fraction with electron ($e$), muon ($\mu$), and tau ($\tau$) decays is respectively 30, 20 and 50 percent.

Backgrounds from top-quark events via pair and electroweak production (top),  {\it V}+jets events, and diboson events ({\it VV}) are all modeled via simulation.    
The {\sc alpgen} generator~\cite{alpgen} is used to estimate  {\it V}+jets (including the ratio of light- to heavy-flavor events), {\sc powheg}~\cite{powheg} for electroweak production of top quarks, and {\sc pythia}~\cite{pythia} for top quark pair production and {\it VV} events, as well as for the {\it VH} signal.
The parton showering is performed by {\sc pythia}. The event generation process includes a simulation of the detector response~\cite{geant}, and the resulting samples are subjected to the same reconstruction and analysis chain as the data. The normalization of the simulated samples is described in Ref.~\cite{tevcomb2012}. 
Electroweak (EWK) mistags, events with light-flavor jets that are wrongly tagged, are mostly due to  {\it V}+jets and are determined from light-flavor simulated samples weighted by a per-event mistag probability, obtained for each algorithm from an orthogonal data sample~\cite{secvtx,jetprob}. 

The background contribution from QCD MJ events is difficult to describe accurately with the simulation, and so is modeled separately from an independent data sample.  We predict the QCD MJ contribution from data events with $\Delta\varphi(\vmet,\vec{E}_\mathrm{T}^{j_2}) <0.4$ and \mbox{$35<\met<70\gev$}. In this sample, we measure the contamination from events with heavy-flavor jets or light-flavor mistags that fall into one of the three previously described tagging categories~\cite{kaThesis}. 

Following Ref.~\cite{Aaltonen:2009jg}, we parametrize this category-tagging rate (the ratio of category-tagged events to events satisfying taggability requirements) in bins of the magnitude of the negative vector sum of the transverse momenta of the charged tracks within the jet ($\vmpt$)~\cite{MPT_NIM}, the scalar sum of transverse energies of $j_1$, $j_2$, and $j_3$ (where applicable) $H_\mathrm{T}$, $\JetPtRelOne$, and $\JetPtRelTwo$, where $\PtRelDef$. We define one 4-dimensional matrix ($M_{\mathit{TR}}$) for each tagging category. The large data sample available allows improvement of this model by defining an \emph{event-based} $M_{\mathit{TR}}$ instead of a \emph{jet-based} $M_{\mathit{TR}}$. The advantage is that correlations between the jets in each event are properly taken into account.
We use the $M_{\mathit{TR}}$ to predict the QCD MJ contribution in the preselection, which has the same flavor composition before tagging requirements as in the region from which the $M_{\mathit{TR}}$ is derived. The QCD MJ background normalization in each tagging category is determined from the corresponding $M_{\mathit{TR}}$ after subtracting the contributions from all other background sources, which are estimated using simulated events. The model is validated in various control regions, defined below.

\begin{figure}
\begin{overpic}[width=\linewidth]{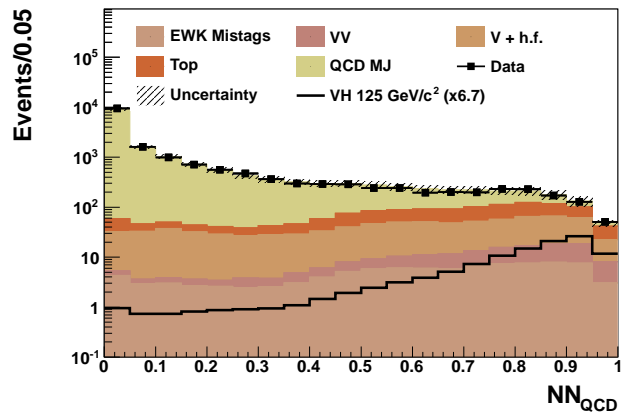}\end{overpic}
\caption{\label{fig:nnqcd}The distribution of ${\rm NN}_{\rm QCD}$ for events satisfying the preselection criteria. The signal region is defined by \mbox{${\rm NN}_{\rm QCD}>0.45$}. The normalization of the QCD MJ contribution is determined from the data. The uncertainty includes all statistical and systematic contributions (see text)}.
\end{figure}

We employ an artificial neural network, ${\rm NN}_{\rm QCD}$, to further separate the dominant QCD MJ background from the signal and other backgrounds. We train a 14-variable feed-forward multilayer perceptron bearing activity-derived (\met, \mpt), angular ($\Delta\varphi(\vmet,\vmpt)$, angular separations between \vmet, \vmpt\ and the jet directions), and event shape (centrality and sphericity~\cite{sphericityTensor}) observables~\cite{kaThesis}. 
Figure~\ref{fig:nnqcd} shows the distribution of ${\rm NN}_{\rm QCD}$ in the preselection sample; the QCD MJ backgrounds (peaking at 0) are well separated from the signal (peaking at 1). 
We retain only events with ${\rm NN}_{\rm QCD}>0.45$ (signal region), rejecting about 90\% (70\%) of the QCD MJ (overall) background while keeping $95\%$ of the signal. This represents a $15\%$ increase in background rejection for the same signal acceptance compared to Ref.~\cite{Aaltonen:2009jg}. The $S/B$ in the signal region is about $1/60$ in the SS tagging category for $m_H=125~\gevcc$, similar to that of the corresponding tagging category of Ref.~\cite{Aaltonen:2009dh}. 

\begin{table}[b]
\caption{\label{tbl:eventYields}Comparison of the number of expected and observed events in the signal region for different tagging categories. The uncertainties include all statistical and systematic contributions (see text)~\cite{supplementalInfoYields}.}
\begin{ruledtabular}
\begin{tabular}{lD{,}{\pm}{6.6}D{,}{\pm}{8.8}}
\toprule
\multirow{2}{*}{Process} & \multicolumn{1}{c}{2 $b$ tags} & \multicolumn{1}{c}{1 $b$ tag} \\
& \multicolumn{1}{c}{SS + SJ} & \multicolumn{1}{c}{ 1S } \\\hline
{\it VV} & 62 , 7.5 & 293 , 32\\
Top & 370 , 52 & 1015 , 128\\
 {\it V} + heavy flavor & 424 , 81 & 3680 , 675\\
EWK mistags &55 , 26 & 2288 , 283\\
QCD MJ & 1300 , 31 & 10825 , 177\\
\hline
Total & 2211 , 197 & 18100 , 1295\\
\hline
Data & \multicolumn{1}{c}{2117} & \multicolumn{1}{c}{18165}\\
\hline\hline
\multicolumn{3}{l}{Expected Higgs boson signal for $m_H=125\gevcc$} \\
\hline
$ZH\to\nu\nu b\bar{b}$ & \multicolumn{1}{c}{7.6}  & \multicolumn{1}{c}{9.7} \\
$WH\to\ell\nu b\bar{b}$ & \multicolumn{1}{c}{8.0} & \multicolumn{1}{c}{10.6} \\
$ZH\to\ell\ell b\bar{b}$ &\multicolumn{1}{c}{0.4} & \multicolumn{1}{c}{0.6} \\
\bottomrule
\end{tabular}
\end{ruledtabular}
\end{table}

We employ a second network, ${\rm NN}_{\rm SIG}$, to discriminate the expected signal from the remaining backgrounds. Seven input variables are used for this purpose: the invariant mass of the two leading jets ($m(j_1,j_2)$), 
the invariant mass of \met\ and all jets, the differences $H_T-\met$ and $\mht-\met$ ($\mht$ is the magnitude of the negative vector sum of jet $E_{\mathrm{T}}$'s), the maximum $\Delta R$ between the jets, the output of ${\rm NN}_{\rm QCD}$, and the output of a NN using tracking information to separate events with intrinsic \met\ from those with instrumental \met~\cite{metbbnn}.

\begin{figure}[!h]
\begin{overpic}[width=.995\linewidth]{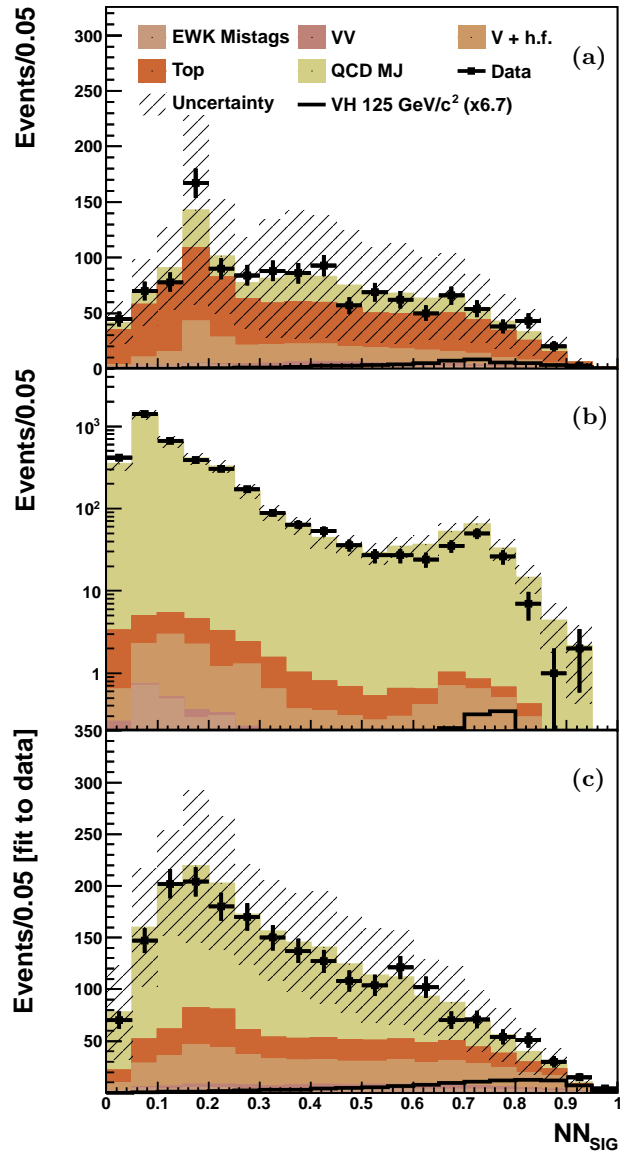}
\put(215,428){\bf(a)}
\put(215,292){\bf(b)}
\put(215,155){\bf(c)}
\end{overpic}
\caption{\label{fig:nnsigCR}The distribution of the final discriminant function, ${\rm NN}_{\rm SIG}$, for events with two $b$ tags (SS+SJ categories) in the control samples: (a) EWK, (b) MJ1, (c) signal region (${\rm NN}_{\rm QCD} > 0.45$). Only the normalization of the QCD MJ is fit to the data.}
\end{figure}

We avoid potential bias by testing our understanding of the SM backgrounds in several control samples where the expected amount of signal is negligible. 
We define an EWK region (Fig.~\ref{fig:nnsigCR}(a)) by requiring events to have at least one charged lepton in addition to satisfying the preselection criteria. This region is sensitive to top-quark pair,  {\it V}+jets, and, to a lesser extent, {\it VV} and electroweak single-top-quark production, and is used to validate the simulation against the data. We also define the MJ1, MJ2, and MJ3 control regions, which contain no identified lepton and are dominated by QCD processes. MJ1 (Fig.~\ref{fig:nnsigCR}(b)) contains events with $\Delta\varphi(\vmet,\vec{E}_\mathrm{T}^{j_{2}}) < 0.4$ and $\met>70\gev$. MJ2 contains events satisfying the preselection requirements and ${\rm NN}_{\rm QCD}<0.1$ and is the region where the QCD MJ normalization is obtained from the data. MJ3, defined from preselection events with $0.1 \leq {\rm NN}_{\rm QCD} \leq 0.45$, serves as a final consistency check of the overall normalization. Finally, we validate our background model in the preselection region before proceeding with the final fit in the signal region.
We check the distribution of multiple kinematic variables, including all inputs to ${\rm NN}_{\rm QCD}$ and to the final discriminant function ${\rm NN}_{\rm SIG}$, defined in the next paragraph, as well as the output of these two networks in all our control samples~\cite{kaThesis}. We obtain good agreement between the data and our SM background model in all the samples, with only the normalization of the QCD MJ component determined from the fit to data. 


The distribution of ${\rm NN}_{\rm SIG}$ is validated in our control samples, as shown in Fig.~\ref{fig:nnsigCR} for events with two $b$ tags. Figure~\ref{fig:nnsigCR}(c) shows the distribution of ${\rm NN}_{\rm SIG}$ in the signal region for events with two $b$ tags. The expected number of events is compared to the observed yields in Table~\ref{tbl:eventYields}.
For $m_H=125\gevcc$, we expect a total of $21$ ($16$) signal events with one (two) $b$-tagged jets. 

We perform a binned likelihood fit to probe for a {\it VH} signal in the presence of SM backgrounds. The likelihood is the product of Poisson probabilities over the bins in the ${\rm NN}_{\rm SIG}$ distribution. The mean number of expected events in each bin includes contributions from each background source and from the {\it VH} processes (assuming a given value of $m_H$).
We employ a Bayesian likelihood method~\cite{limits} with a flat, non-negative, prior probability for the SM Higgs boson production cross section times branching fraction, \mbox{$\sigma(VH)\times\mathcal{B}(H\to b\bar{b})$}, and truncated Gaussian priors for the uncertainties on the acceptance and shape of the backgrounds. We combine the three tagging categories by taking the product of their likelihoods and simultaneously varying the correlated uncertainties. All systematic uncertainties except those associated with the QCD MJ and the EWK mistags are treated as fully correlated across the tagging categories. 
 
The uncertainties from the simulations statistics and those on the normalizations of top-quark (10\%), diboson (6\%),  {\it V}+jets (30\%), QCD MJ (1 to 3\%), and EWK mistags (20 to 65\%) production are not correlated.
The shapes obtained by varying the $M_{\mathit{TR}}$ (mistag) probabilities by one standard deviation from their central
values are applied as shape uncertainties for the QCD MJ (EWK mistags).
The correlated uncertainties, which apply to both the signal and the EWK backgrounds, include luminosity measurement (6\%), $b$-tagging efficiency (5 to 10\%), trigger efficiency (3-5\%), lepton veto efficiency (2\%), parton distribution function (3\%), and up to 11\% for the jet-energy scale~\cite{cdfjets}. We also determine the shape uncertainties on ${\rm NN}_{\rm SIG}$ due to the jet-energy scale and the trigger efficiency. The latter two also affect the QCD MJ background through the background subtraction procedure described above. Initial- and final-state radiation uncertainties (2 to 3\%) are applied only to the {\it VH} signal. 

We compute 95\% C.L. upper limits on \mbox{$\sigma(VH)\times\mathcal{B}(H\to b\bar{b})$} for \mbox{$90 < m_H <150\gevcc$} in $5\gevcc$ steps using the methodology described in Ref.~\cite{Aaltonen:2010jr}. The expected and observed upper limits are shown in Table~\ref{tbl:limits}. 
We test the consistency of the observed limits with the signal hypothesis by statistical sampling of the signal-plus-background model (assuming $m_H=125~\gevcc$). These studies indicate that the median upper C.L. in the SM Higgs scenario is higher (up to $2.5$ units in SM cross-section) than that of the background-only hypothesis over the $90-150~\gevcc$ range, and is consistent with the observed limits within one standard deviation.

In summary, we have performed a direct search for the SM Higgs boson decaying into $b\bar{b}$ pairs using the full CDF~II data sample, corresponding to 9.45\invfb\ of integrated luminosity accumulated during Run II of the Tevatron. Improved techniques increase the sensitivity by roughly 15\% with respect to a previous analysis~\cite{Aaltonen:2009jg} in addition to the improvement due to larger integrated luminosity.
We set 95\% C.L. upper limits on \mbox{$\sigma(VH)\times\mathcal{B}(H\to b\bar{b})$} for \mbox{$90 < m_H <150\gevcc$} with $5\gevcc$ increments. 
For a Higgs boson mass of  $125\gevcc$, the observed limit is $6.7$ times the SM prediction, consistent with the expected limit of $3.6$ within two standard deviations.  

\begin{figure}[tb]
\includegraphics[width=\linewidth]{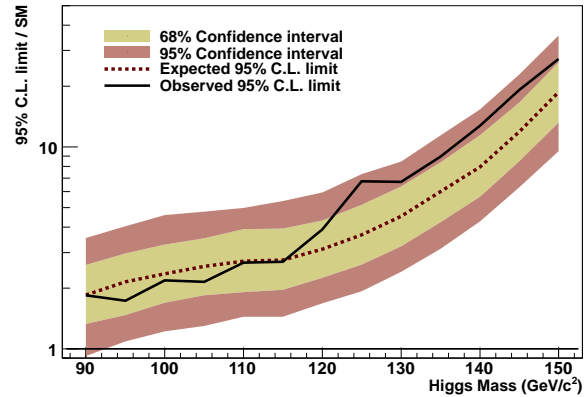}
\caption{\label{fig:limitPlot}
Observed and expected (median, for the background-only hypothesis) 95\% C.L. upper limits on {\it VH} cross section times $\mathcal{B}(H\to b\bar{b})$ divided by the  SM prediction, as a function of the Higgs boson mass. The bands indicate the 68\% and 95\% credibility regions where the limits can fluctuate, in the absence of signal. 
}
\end{figure}

\begin{table}[t]
\caption{\label{tbl:limits}Expected and observed 95\% C.L. upper limits on the {\it VH} cross section times $\mathcal{B}(H\to b\bar{b})$ and their ratio to the SM prediction~\cite{tevcomb2012}.}
\begin{ruledtabular}
\begin{tabular}{l*{2}{D{.}{.}{4.7}D{.}{.}{-1}}}
$m_H$ & \multicolumn{2}{c}{$\sigma_{VH} \times \mathcal{B}(H\to b\bar{b})$ (pb)} & \multicolumn{2}{c}{Ratio to SM prediction}\\
($\mathrm{GeV}/c^2$) & \multicolumn{1}{c}{Expected} & \multicolumn{1}{c}{Observed} & \multicolumn{1}{c}{Expected} & \multicolumn{1}{c}{Observed}  \\\hline
90 & 0.92^{+0.40}_{-0.27} & 0.92 & 1.8 ^{+0.8}_{-0.5} & 1.8 \\
95 & 0.91^{+0.34}_{-0.29} & 0.73 & 2.2 ^{+0.8}_{-0.7} & 1.7 \\
100 & 0.82^{+0.33}_{-0.24} & 0.77 & 2.3 ^{+0.9}_{-0.7} & 2.2 \\
105 & 0.75^{+0.30}_{-0.21} & 0.63 & 2.6 ^{+1.0}_{-0.7} & 2.1 \\
110 & 0.65^{+0.28}_{-0.19} & 0.64 & 2.7 ^{+1.2}_{-0.8} & 2.8 \\
115 & 0.54^{+0.23}_{-0.16} & 0.53 & 2.7 ^{+1.2}_{-0.8} & 2.7 \\
120 & 0.49^{+0.20}_{-0.13} & 0.61 & 3.1 ^{+1.3}_{-0.9} & 3.9 \\
125 & 0.44^{+0.17}_{-0.12} & 0.81 & 3.6 ^{+1.4}_{-1.0} & 6.7 \\
130 & 0.41^{+0.17}_{-0.12} & 0.60 & 4.6 ^{+1.9}_{-1.4} & 6.7 \\
135 & 0.38^{+0.16}_{-0.11} & 0.57 & 6.0 ^{+2.5}_{-1.8} & 8.9 \\
140 & 0.34^{+0.15}_{-0.10} & 0.55 & 8.0 ^{+3.4}_{-2.3} & 12.7 \\
145 & 0.33^{+0.13}_{-0.09} & 0.53 & 11.8 ^{+4.8}_{-3.4} & 19.2 \\
150 & 0.30^{+0.13}_{-0.09} & 0.45 & 18.4 ^{+7.6}_{-5.2} & 27.2 \\
\end{tabular}
\end{ruledtabular}
\end{table}

We thank the Fermilab staff and the technical staffs of the
participating institutions for their vital contributions. This work
was supported by the U.S. Department of Energy and National Science
Foundation; the Italian Istituto Nazionale di Fisica Nucleare; the
Ministry of Education, Culture, Sports, Science and Technology of
Japan; the Natural Sciences and Engineering Research Council of
Canada; the National Science Council of the Republic of China; the
Swiss National Science Foundation; the A.P. Sloan Foundation; the
Bundesministerium f\"ur Bildung und Forschung, Germany; the Korean
World Class University Program, the National Research Foundation of
Korea; the Science and Technology Facilities Council and the Royal
Society, UK; the Russian Foundation for Basic Research; the Ministerio
de Ciencia e Innovaci\'{o}n, and Programa Consolider-Ingenio 2010,
Spain; the Slovak R\&D Agency; the Academy of Finland; and the
Australian Research Council (ARC).

\bibliography{apssamp}

\begin{thebibliography}{99}
\bibitem{coord} CDF uses a cylindrical coordinate system with the $z$ axis along the proton beam axis. The pseudorapidity is \mbox{$\eta=-\ln(\tan(\frac{\theta}{2}))$}, where $\theta$ is the polar angle, and $\varphi$ is the azimuthal angle, 
while \mbox{$p_{\mathrm{T}}=p \sin\theta$} and \mbox{$E_{\mathrm{T}}=E \sin\theta$}. The $\met$ is defined as the magnitude of $\vmet = -\sum_{i} E_{\mathrm{T}}^i\hat{n}_i$, where $\hat{n}_i$ is a unit vector perpendicular to the beam axis and pointing at the $i$th calorimeter tower, and $E_{\mathrm{T}}^i$ is the transverse energy therein.
\bibitem{triggerPath} A trigger path is uniquely defined by specifying selection criteria at each of the three trigger levels. The trigger paths and the parametrization of their combined efficiency are described in detail in Ref.~\cite{kaThesis}.
\bibitem{sphericityTensor}
The event sphericity is defined by $S=\frac{3}{2}(\lambda_2+\lambda_3)$, where the sphericity tensor is \mbox{$S^{\alpha\beta} = (\sum_i p_i^\alpha p_i^\beta) / (\sum_i p_i^2)$} and $\lambda_1>\lambda_2>\lambda_3$ are its three eigenvalues and satisfy $\lambda_1+\lambda_2+\lambda_3=1$. The index $i$ refers to each jet in the event. The event centrality is \mbox{$C=\frac{\sqrt{(\sum_i p^i_x)^2+(\sum_i p^i_y)^2}}{\sqrt{(\sum_i p^i_x)^2+(\sum_i p^i_y)^2+(\sum_i p^i_z)^2}}$}.
\end{thebibliography}

\end{document}